\documentclass{ws-procs975x65}
\usepackage{amsmath}

\def\E{\textbf{E}\ }
\def\B{\textbf{B}\ }
\def\F{\textbf{F}\ }
\def\J{\textbf{J}\ }
\def\RE{\textbf{R} } 
\def\R{R } 
\def\PR{\mathcal{R}} 

\newcommand{\M}{{\bf M}}   

\newcommand{\MM}{{\cal M}} 

\newcommand{\MX}{{\cal X}}
\newcommand{\MY}{{\cal Y}}

\def\R{\ensuremath{\mathcal{R}}}   
\def\ihat{\^{i}}
\def\jhat{\^{j}}
\def\khat{\^{k}}

\begin{document}



\title{Time machines and quantum theory}

\author{Mark J Hadley}

\address{Department of Physics,\\
 University of Warwick, \\
 Coventry\\
CV4~7AL\\ UK\\
\email{Mark.Hadley@warwick.ac.uk}}


%

\begin{abstract}
There is a deep structural link between acausal spacetimes and quantum theory. As a consequence quantum theory may resolve some "paradoxes" of time travel. Conversely, non-time-orientable spacetimes naturally give rise to electric charges and spin half. If an explanation of quantum theory is possible, then general relativity with time travel could be it.
\end{abstract}
\bodymatter

\section{Introduction}
Wormholes, Time machines and Energy conditions: would Einstein have come to this parallel session? If it was held in 1950, I don't think so. Einstein had a strong belief in preserving causality as a fundamental principle that contributed to his work on relativity. But if Einstein was at MG11 in 2006, this is where Einstein would be. Einstein would be attracted not so much for his interest in general relativity but for his desire to find an explanation for quantum theory. The difference between the two dates is our understanding of quantum theory, particularly Bell's inequalities which showed that quantum theory could not be explained by any local hidden variables theory. Or in other words; if an underlying explanation for quantum theory is ever to be found it must involve time travel in a technical sense. The intervening years have also seen experimental verification of the violation of Bell's inequalities. Unlike Einstein's EPR thought experiment, the conclusion cannot be that the equations are deficient, but that quantum theory is either inexplicable or requires a non local unifying theory.

Perhaps quantum theory is telling us something about spacetime. If so, where better to look for a deeper understanding than the best theory of space and time that we have - general relativity. In fact, it is our only theory of space and time, so the choice of where to start is obvious!

We already know:
\begin{itemize}
  \item That solutions of General Relativity with closed timelike curves exist (mathematically)
  \item That there is an intimate link between acausal spacetimes and quantum logic \cite{hadley97}
  \item That the paradoxes of time travel are incorrect solutions.
\end{itemize}

A whole range of acausal solutions to the equations of general relativity exist from simple flat tori models, the Godel Universe and wormholes. The possibility of a traversable wormhole remains a tantalising proposition. Being able to create an acausal region from a simple spacetime is heavily constrained by theorems such as Geroch \cite{geroch} and Tipler \cite{tipler} but speculations about a chronology protection mechanism have not been fruitful. The main constraint on traversable wormholes is the requirement is some form of negative energy. Research activity over the years has focussed on the most likely candidates. Casmir energy was one credible candidate, more recently the growing cosmological evidence for \emph{dark energy} has prompted models with the so called \emph{phantom energy} - a prominent theme in this session.

The logical and structural links between quantum theory and  general relativity with acausal structure are known but not widely appreciated. This is partly because the logical structure of quantum theory is not well known, neither is its significance - these are discussed later in the paper.

Time travel is enigmatic, most science fiction examples involve, or invoke, implicit paradoxes where a signal from the future creates an event that was not in its past. It is the paradoxes of macroscopic objects travelling back in time that led to a chronology protection conjecture \cite{hawking}. In a mathematical and physical sense there are no paradoxes, because when they are examined they are simply inconsistent solutions to the problem, or to be more precise they are not solutions. As an analogy, consider a standing wave in a string fixed at both ends; it has a precise frequency that forms a valid solution to the equations and boundary conditions. If the frequency is changed slightly, the wave starting at one end will reach the far end with a non-zero amplitude apparently moving a fixed end point. Is this a paradox? No, it is just a wrong answer. The correct answer does not allow arbitrary frequencies. Similarly for time travel, consistency imposes constraints on allowed solutions.

It has been suggested that self-consistency arises from the application of an action principle in a natural way \cite{carlini}. Certainly some addition to or alternative to the classical equations is required to ensure consistency and remove the so called paradoxes. I will argue that the new rules required to adapt classical physics to acausal spacetimes is none other than quantum theory.

\section{The Essence of quantum theory}

Quantum theory has many strange features, but most are not absolutely unique to quantum theory. For example the uncertainty relationship between position and momentum is common to classical waves, or the wave equations themselves. While some, apparently critical, features are largely irrelevant; such as the inability to make simultaneous measurements of some observables - if this were possible  (even for non-commuting observables) it would not change quantum theory.

A fundamental difference between quantum and classical physics can be found in the logical structure of the propositions. Classical physics satisfies Boolean logic while quantum theory is non-Boolean. The distributive law does not hold, instead the propositions form an orthomodular lattice \cite {beltrametti_cassinelli}. This leads to the requirement to represent probabilities as subspaces of a [complex] vector space rather than as measures of a volume space. It is this fundamental distinction that leads, with some symmetry arguments, to the dynamical and statistical features of quantum theory.

 Although the relationship between incompatible observables has a non-Boolean characteristic, in any one experiment the propositions satisfy the normal Boolean logic. This is generally expressed by saying that quantum theory is {\em context dependent} - within any single context, the probabilities can be expressed as usual. In a formal way, the propositions satisfy an orthomodular lattice, which is a weaker condition that a Boolean or distributive lattice.

  In all of classical physics probability can be ascribed to our ignorance of some variables (typically the precise initial conditions). In quantum theory this interpretation seems not to be possible. Certainly, it cannot be described in terms of local hidden variables - as shown by the violations of Bell's inequalities. If an explanation is possible then it must have non-local features and the usual concept of probability could be restored. Probabilities would arise from our ignorance of initial or boundary conditions, but some of these conditions would be in the experimenter's future. If an explanation for quantum theory is not possible, then we have to accept a completely new concept of probability, where uncertainty is not due to ignorance, but is a fundamental feature of Nature.
%

A formal proof that acausal spacetimes lead to the logical structure of quantum theory is given in the paper by Hadley \cite{hadley97}. The proof is formal and can be largely replaced by the statement that an acausal spacetime is context dependent. This can been seen simply in two ways:

With closed timelike curves (CTCs) or a failure of time orientability, it is not in general possible to set up boundary conditions on an {\em initial} surface without some knowledge of {\em future} conditions. Future experiments can set extra boundary conditions that are not redundant. Simple models with Billiards such as those described by Carlini and Novikov \cite{carlini} provide a good illustration. The standing waves on a string provide a helpful analogy, boundary conditions need to be specified at both ends before the wave can be calculated.

An alternative way to look at the context dependence, with the same example, would be to consider the shape of the standing wave as a combination of forward and backward moving waves. A change of time direction  be like at a future experiment would {\em send a signal back} to the start of the experiment. This can be seen as a realisation of the Cramer's transactional interpretation of quantum theory \cite{cramer}.

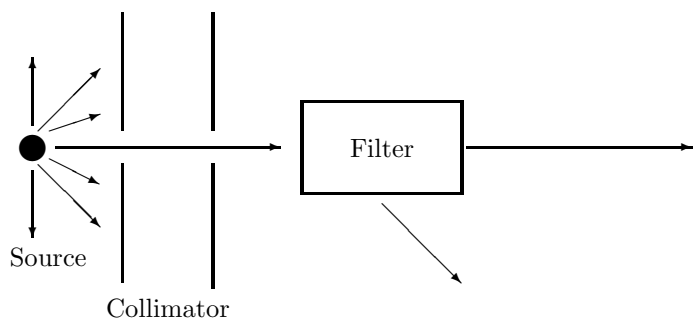
\begin{figure}[ht]

\setlength{\unitlength}{.01mm}

\begin{picture}(12876,5452)(-599,-5191) 

\thicklines
\put(600,-2161){\circle*{336}}
\put(1800,-361){\line( 0,-1){1575}}
\put(3000,-361){\line( 0,-1){1575}}
\put(1800,-2380){\line( 0,-1){1575}}
\put(3000,-2380){\line( 0,-1){1650}}
\put(4200,-2761){\framebox(2100,1200){Filter}}

\thinlines
\put(5251,-2911){\vector( 1,-1){1050}}
\put(826,-1936){\vector( 3, 1){675}}
\put(826,-2311){\vector( 2,-1){660}}
\put(676,-2386){\vector( 1,-1){825}}
\put(676,-1936){\vector( 1, 1){825}}
\put(600,-1861){\vector( 0, 1){900}}
\put(600,-2461){\vector( 0,-1){900}}
\put(900,-2161){\vector( 1, 0){3000}}
\put(6376,-2161){\vector( 1, 0){3000}}
\put(300,-3736){\makebox(0,0)[lb]{Source}}
\put(2400,-4411){\makebox(0,0)[cb]{Collimator}}
\end{picture}
\caption{The state-preparation}
\label{fig:init}
\end{figure}

We can see how the non-Boolean logic arises and how the normal probability rules appear to be violated by considering the measurement of the spin of a spin-half particle in two different orientations. In general you should be wary of examples where the results form a two dimensional set ( eg spin up or down) - because some general results about quantum theory, such as Gleason's theorem, admit counter examples in two dimensions. However the arguments presented here can easily be extended to higher dimensional spaces. We can see the relationship between causality and probability structures by considering the spacetime manifolds that correspond to different results from a variety of experiments. The classical view would be that the spacetime is an evolving 3-manifold. Probabilities could be evaluated by putting a measure on all possible initial 3-manifolds and calculating the fraction that evolve to each experimental outcome. Applying this approach, let us constrain the possible manifolds by setting up the state preparation apparatus as depicted in Fig.~\ref{fig:init}.

We denote the set of all 3-manifolds consistent with the state preparation by, $\MM$. We can then measure the x-spin or y-spin with different experimental set ups. Each measurement will divide up $\MM$ into two parts; those giving a positive or spin-up result and those giving a negative or spin-down result, we denote these subsets as $\MX^+$ and $\MX^-$ respectively. In general the two parts will be unequal and non-zero corresponding to a state preparation that gives different probabilities to the two outcomes. We can divide $\MM$ similarly according to the results of a y-measurement into $\MY^+$ and $\MY^-$ as shown by the Venn diagram in Fig.~\ref{fig:classicalvenn}. Classically, the probabilities would be proportional to the weighted areas.

\begin{figure}[ht]

\setlength{\unitlength}{.01mm}

\begin{picture}(12876,5452)(-599,-5191) 

\thicklines
\put(12000,239){\line( 0,-1){4800}}
\put(11700,239){\line( 1, 0){300}}
\put(11700,-61){\line( 1, 0){300}}
\put(11700,-361){\line( 1, 0){300}}
\put(11700,-661){\line( 1, 0){300}}
\put(11700,-961){\line( 1, 0){300}}
\put(11700,-1261){\line( 1, 0){300}}
\put(11700,-1561){\line( 1, 0){300}}
\put(11700,-1861){\line( 1, 0){300}}
\put(11700,-2161){\line( 1, 0){300}}
\put(11700,-2461){\line( 1, 0){300}}
\put(11700,-2761){\line( 1, 0){300}}
\put(11700,-3361){\line( 1, 0){300}}
\put(11700,-3661){\line( 1, 0){300}}
\put(11700,-3961){\line( 1, 0){300}}
\put(11700,-4261){\line( 1, 0){300}}
\put(11700,-4561){\line( 1, 0){300}}
\put(11700,-3061){\line( 1, 0){300}}
\put(11700,-4861){\makebox(0,0)[b]{$x$-position}}
\put(11700,-5161){\makebox(0,0)[b]{measurement}}
\put(7200,-2761){\framebox(2100,1200){}}
\put(8250,-1861){\makebox(0,0)[b]{\small $x$-oriented}}
\put(8250,-2161){\makebox(0,0)[b]{\small Stern-}}
\put(8250,-2461){\makebox(0,0)[b]{\small Gerlach}}
\put(600,-2161){\circle*{336}}
\put(1800,-361){\line( 0,-1){1575}}
\put(3000,-361){\line( 0,-1){1575}}
\put(1800,-2380){\line( 0,-1){1575}}
\put(3000,-2380){\line( 0,-1){1650}}
\put(4200,-2761){\framebox(2100,1200){Filter}}

\put(8100,-4861){\vector( 0, 1){600}}
\put(8100,-4861){\vector( 1, 0){600}}

\thinlines
\put(5251,-2911){\vector( 1,-1){1050}}
\put(826,-1936){\vector( 3, 1){675}}
\put(826,-2311){\vector( 2,-1){660}}
\put(676,-2386){\vector( 1,-1){825}}
\put(676,-1936){\vector( 1, 1){825}}
\put(600,-1861){\vector( 0, 1){900}}
\put(600,-2461){\vector( 0,-1){900}}
\put(900,-2161){\vector( 1, 0){3000}}
\put(9600,-2461){\vector( 3,-2){1453.846}}
\put(9600,-1861){\vector( 3, 2){1453.846}}
\put(6376,-2161){\vector( 1, 0){675}}
\put(300,-3736){\makebox(0,0)[lb]{Source}}
\put(2400,-4411){\makebox(0,0)[cb]{Collimator}}
\put(8400,-5161){\makebox(0,0)[b]{$z$}}
\put(7800,-4561){\makebox(0,0)[b]{$x$}}
\end{picture}
\caption{State-preparation and an $x$-spin measurement}
\label{fig:xmeas}
\end{figure}
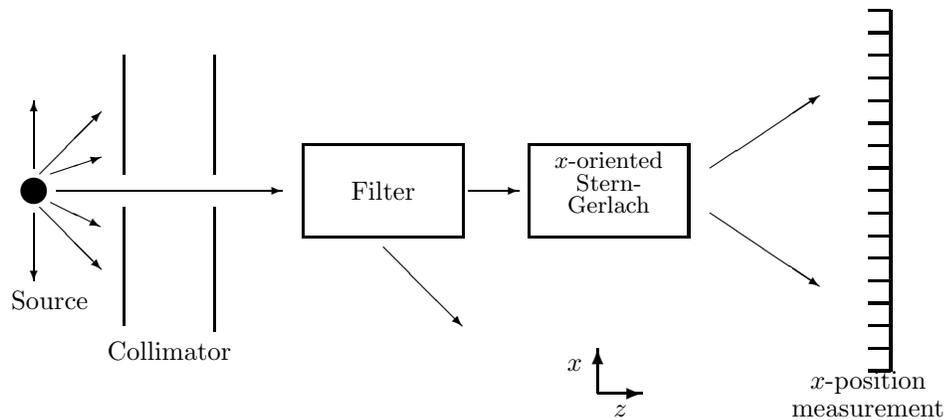

\begin{figure}[ht]
\setlength{\unitlength}{.01mm}

\begin{picture}(12876,5452)(-599,-5191) 
\thicklines
\put(7200,-2761){\framebox(2100,1200){}}
\put(600,-2161){\circle*{336}}
\put(1800,-361){\line( 0,-1){1575}}
\put(3000,-361){\line( 0,-1){1575}}
\put(1800,-2380){\line( 0,-1){1575}}
\put(3000,-2380){\line( 0,-1){1650}}
\put(4200,-2761){\framebox(2100,1200){Filter}}

\put(11700,-3961){\line( 1, 0){600}}
\put(10200,-1561){\line( 1, 0){375}}
\put(10426,-1861){\line( 1, 0){375}}
\put(9889,-656){\line( 3,-4){2448}}
\put(11551,-3661){\line( 1, 0){525}}
\put(11400,-3361){\line( 1, 0){450}}
\put(11251,-3061){\line( 1, 0){375}}
\put(11026,-2761){\line( 1, 0){375}}
\put(10876,-2461){\line( 1, 0){300}}
\put(10651,-2161){\line( 1, 0){300}}
\put(9600,-661){\line( 1, 0){300}}
\put(9826,-961){\line( 1, 0){300}}
\put(9976,-1261){\line( 1, 0){375}}

\put(11026,-2161){\vector( 0, 1){600}}
\put(11026,-2161){\vector( 1, 0){600}}
\put(11026,-2176){\vector( 3,-4){340}}

\thinlines
\put(5251,-2911){\vector( 1,-1){1050}}
\put(826,-1936){\vector( 3, 1){675}}
\put(826,-2311){\vector( 2,-1){660}}
\put(676,-2386){\vector( 1,-1){825}}
\put(676,-1936){\vector( 1, 1){825}}
\put(600,-1861){\vector( 0, 1){900}}
\put(600,-2461){\vector( 0,-1){900}}
\put(900,-2161){\vector( 1, 0){3000}}
\put(6376,-2161){\vector( 1, 0){675}}

\put(9500,-1951){\vector( 1, 1){400}}
\put(9500,-2311){\vector( 3,-2){1400}}

\put(300,-3736){\makebox(0,0)[lb]{Source}}
\put(2400,-4411){\makebox(0,0)[cb]{Collimator}}

\put(7200,-2761){\framebox(2100,1200){}}
\put(8250,-1861){\makebox(0,0)[b]{\small $y$-oriented}}
\put(8250,-2161){\makebox(0,0)[b]{\small Stern-}}
\put(8250,-2461){\makebox(0,0)[b]{\small Gerlach}}

\put(11851,-2161){\makebox(0,0)[b]{$z$}}
\put(11251,-1636){\makebox(0,0)[b]{$x$}}
\put(11476,-2566){\makebox(0,0)[lb]{$y$}}
\put(11896,-4396){\makebox(0,0)[b]{$y$-position}}
\put(11911,-4681){\makebox(0,0)[b]{measurement}}

\end{picture}
\caption{State-preparation and a $y$-spin measurement}
\label{fig:ymeas}
\end{figure}
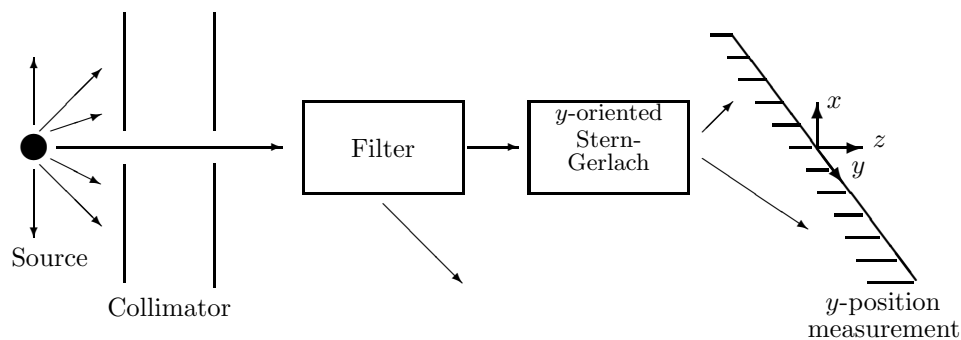

\setlength{\unitlength}{0.8mm}
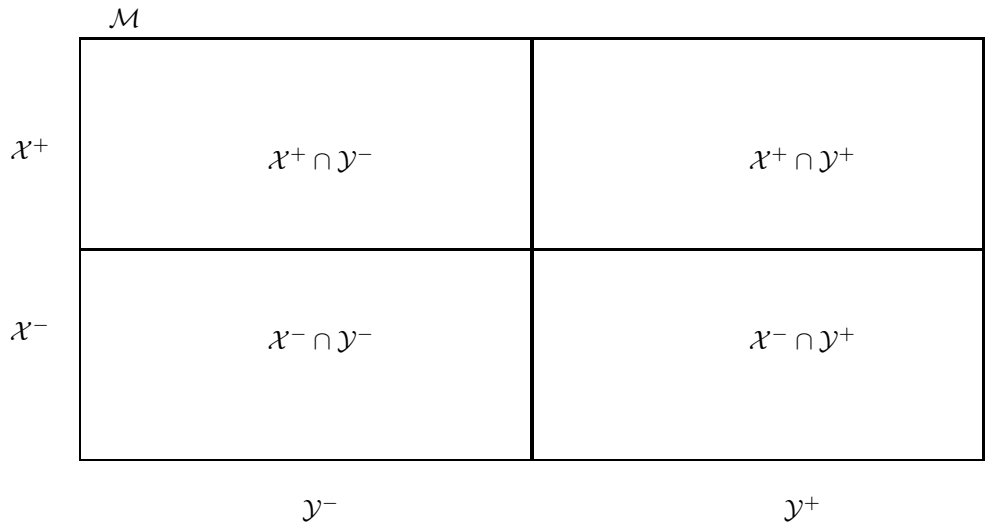
\begin{figure}[ht]
\begin{picture}(220,100)(15,0)
\put(10,10){\framebox(150,70)[tl]{}}
\put(20,82){\makebox(0,0)[br]{$\MM$}}

\thicklines

\put(5,60){\makebox(0,0)[br]{$\MX^+$}}
\put(10,45){\line(1,0){150} }
\put(5,30){\makebox(0,0)[br]{$\MX^-$}}

\put(85,10){\line(0,1){70} }
\put(50,0){\makebox(0,0)[b]{${\bf \MY^-} $}}
\put(130,0){\makebox(0,0)[b]{${\bf \MY^+} $}}

\put(130,60){\makebox(0,0)[c]{$\MX^+ \cap \MY^+ $}}
\put(50,60){\makebox(0,0)[c]{$\MX^+ \cap \MY^- $}}
\put(130,30){\makebox(0,0)[c]{$\MX^- \cap \MY^+ $}}
\put(50,30){\makebox(0,0)[c]{$\MX^- \cap \MY^- $}}

\thinlines
\end{picture}
\caption{A classical view showing sets of 3-manifolds corresponding to different measurement results}
\label{fig:classicalvenn}
\end{figure}

However the classical view of evolving 3-manifolds, is a type of local hidden variable model. It is contradicted by quantum theory and experiment. In our simple example the Venn diagram shows the set of all possible 3-manifold divided into four different sets each of which gives definite values for both the x-spin and y-spin. Quantum theory cannot describe such a state and experimentally it cannot be prepared.

If we are dealing with an acausal spacetime then the preceding arguments require a modification because the possible solutions cannot simply be described as evolving 3-manifolds. Not only the state preparation but also the measurement process itself can set non-redundant boundary conditions and further restrict the set $\MM$ of possible manifolds. Given a full experimental arrangement, the possible results can still be described as sets and subsets of manifolds in a classical way.

\setlength{\unitlength}{0.8mm}
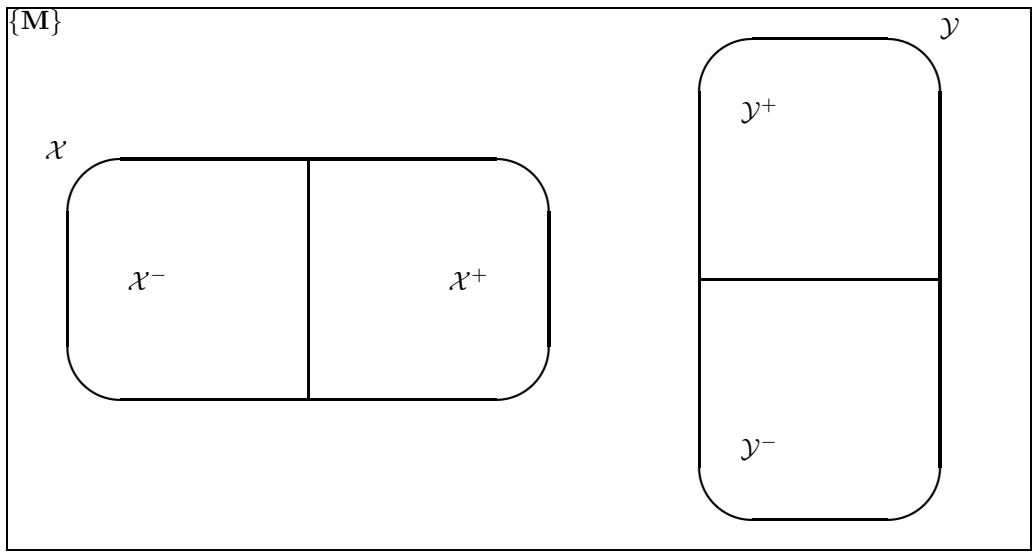
\begin{figure}[ht]
\begin{picture}(220,100)(15,0)
\put(5,5){\framebox(170,90)[tl]{$\{\M\}$}}

\thicklines

\put(55,50){\oval(80,40)}
\put(15,70){\makebox(0,0)[br]{$\MX$}}
\put(55,30){\line(0,1){40} }
\put(25,50){\makebox(0,0)[l]{$\MX^-$}}
\put(85,50){\makebox(0,0)[r]{$\MX^+$}}

\put(140,50){\oval(40,80)}
\put(160,90){\makebox(0,0)[bl]{$\MY$}}
\put(120,50){\line(1,0){40} }
\put(130,20){\makebox(0,0)[b]{${\bf \MY^-} $}}
\put(130,80){\makebox(0,0)[t]{${\bf \MY^+} $}}

\thinlines
\end{picture}
\caption{Sets of 4-manifolds consistent with both state preparation
and the boundary conditions imposed by different measurement
conditions.}
\label{fig:venn}
\end{figure}

A diagrammatic representation of the manifolds is shown in figures~\ref{fig:venn}. As before $\MM$ is the set of all manifolds consistent with state preparations, but now there are subsets $\MX$ and $\MY$ for all manifolds consistent with state preparation and a particular measurement apparatus. The x and y measurements are achieved with incompatible experimental arrangements and therefore set incompatible boundary conditions. Consequently the sets $\MX$ and $\MY$ are disjoint and $\MX^+ \cap \MY^+ = \emptyset$. A result compatible with quantum theory and experiment.

The Venn diagrams give neat illustrations, but also have a greater significance because of the relationship to the propositions (the yes-no questions that apply to the experiments) and the propositional logic. Acausal spacetimes and quantum theory satisfy an orthomodular lattice of propositions \cite{hadley97}. While they break the distributive law for the meet and join of proposition (represented by the Union and intersection of subsets), the essence of the difference lies in how negation is treated. In the classical case the negation of a proposition would correspond to the set $\MM \cap \neg \MX^+$ while in the non-classical case the negation would be in relation to a particular experiment $\MX \cap \neg \MX^+$.

The propositional logic is more than a mathematical curiosity, it is the essence of quantum theory. The only known way to represent probabilities on an orthomodular lattice of propositions is using spaces and subspaces of a Hilbert space \cite{beltrametti_cassinelli}. From a complex Hilbert space, with assumptions about continuity and symmetry of spacetime, the normal equations of quantum theory can be derived see Ballentine~\refcite{ballentine} for the non-relativistic case, and Weinberg~\cite{weinberg95} for the relativistic case.

A particle-like topological structure in space was described by Wheeler, who used the term geon. with the description given above it would be a 3-manifold evolving with time. Extending the idea to a particle-like structure with non-trivial causal structure, for which Hadley\cite{hadley97} has used the term 4-geon. A 4-manifold describes both the particle and its evolution; for a 4-geon they are inseparable. Consequently, the terms {\em initial\/} and {\em evolution\/} need to be used with great care. Although valid in the asymptotically flat region (and hence to any observer), they cannot be extended throughout the manifold. Preparation {\em followed by\/} measurement is also a concept valid only in the asymptotic region: {\em within the particle causal structure breaks down}.

\section{Interactions and topology change}

If particles are modeled as topological structures of space, then interactions such as annihilation of particles, particle-antiparticle creation and some scattering experiments will require the topology of a region of spacetime to change. However there are powerful theorems due to Geroch \cite{geroch} and Tipler \cite{tipler} that constrain topology change in classical general relativity. Topology change in classical general relativity  would need to be a counterexample to Geroch's theorem. The possible counterexamples can be classified as\cite{hadley98}:

\begin{enumerate}
\item Singularities
\item Closed timelike curves
\item A lack of time orientation
\end{enumerate}

The first is a breakdown of general relativity. Since General Relativity is expressed in terms of a 4D spacetime manifold, a singularity is not consistent with a description s spacetime as a manifold. Nevertheless some authors (eg Sorkin) \cite{sorkin97} have proposed a reformulation of general relativity to alow singularities where topology change takes place.

For closed time like curves to appear in a region of space that was previously regular requires negative energy due to Tipler's theorem. While this cannot be ruled out absolutely, with current theoretical knowledge it is unwelcome as a postulate.

The third option is that some timelines from the initial surface turn around and return through the same surface. This would be a failure of time-orientability. Further work on structures that lack time orientability have proved fascinating and fruitful.

Manifolds that do not admit a time orientation can be constructed in a number of ways. The Einstein Rosen bridge is the earliest example, although its non-orientability is often overlooked or misunderstood. The common description of a path through the bridge is that the traveller goes into a black hole and comes out of a white hole. In fact both ends of the wormhole structure are black holes, but the traveller has his time direction reversed, making all black holes look like white holes.

Wormholes can be constructed with any combination of time and space orientability\cite{visser}, and by a similar process monopole type structures of any orientation can be described\cite{diemer_hadley}. The M\"{o}bius strip is non-orientable and if the central circle is considered to be a space dimension, and the other direct a time direction, then the M\"{o}bius strip models a $1+1$ dimensional spacetime that is space orientable but not time orientable.

\section{Tests of orientability}
The orientability of space and time are unambiguous mathematical properties of a manifold. A failure of orientability exists if there is a closed curve (closed in space and time) around which the orientation changes, so that it cannot be globally defined along the curve. However, for a physicist time orientability is neither simple nor unambiguous. A test of space orientability is illustrated and the problems extending the test to time-orientability are described.

Consider a space that is flat and space-orientable everywhere except
for a region \R. The orientation of \R\ is tested by sending a probe into \R. The probe
contains a triad of unit vectors \ihat, \jhat and \khat.  The triad
defines a space-orientation continuously along the path taken by the
probe. The path (world line) of the probe and the path (world line) of the
observer together define a closed loop. An orientation can be defined
in the laboratory, excluding region \R. If the probe that emerges from
\R\ has a different orientation from that defined in the rest of the
laboratory then space is not orientable (see Fig.~\ref{fig:space}).

\setlength{\unitlength}{1mm} \input epsf
\begin{figure}[ht]
\begin{picture}(150,200)
\put(15,-100){\makebox{\epsfbox{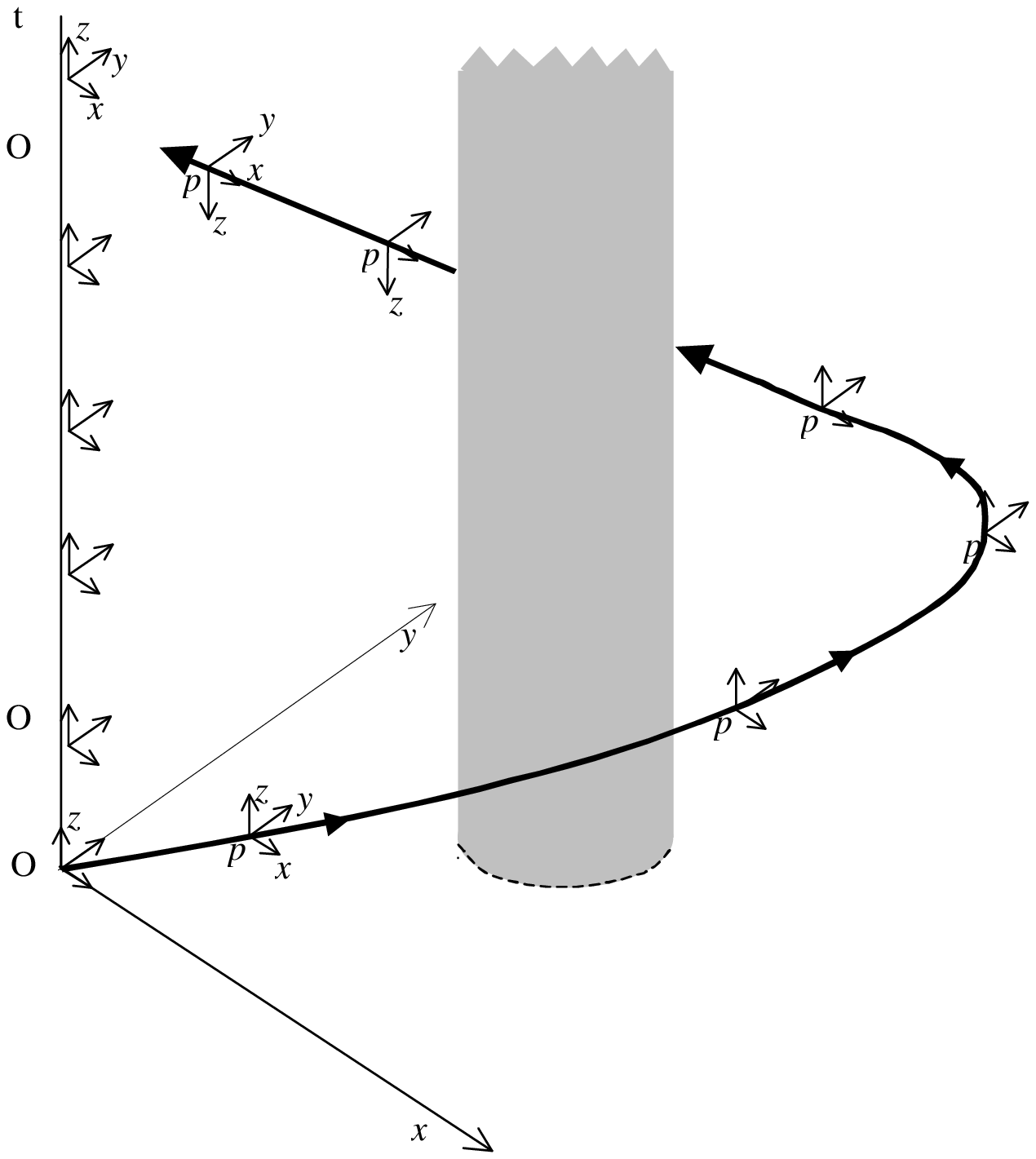}}   }
\put(98,125){\makebox{\R} }
\end{picture}
\caption{A spacetime  diagram showing a positive test  that region \R\
is not time-orientable.  Observer, O,  sends a probe, P, through \R, it
exits \R\ and returns to O with opposite orientation.}
\label{fig:space}
\end{figure}

\begin{figure}[ht]
\begin{picture}(150,200)
\put(15,-100){\makebox{\epsfbox{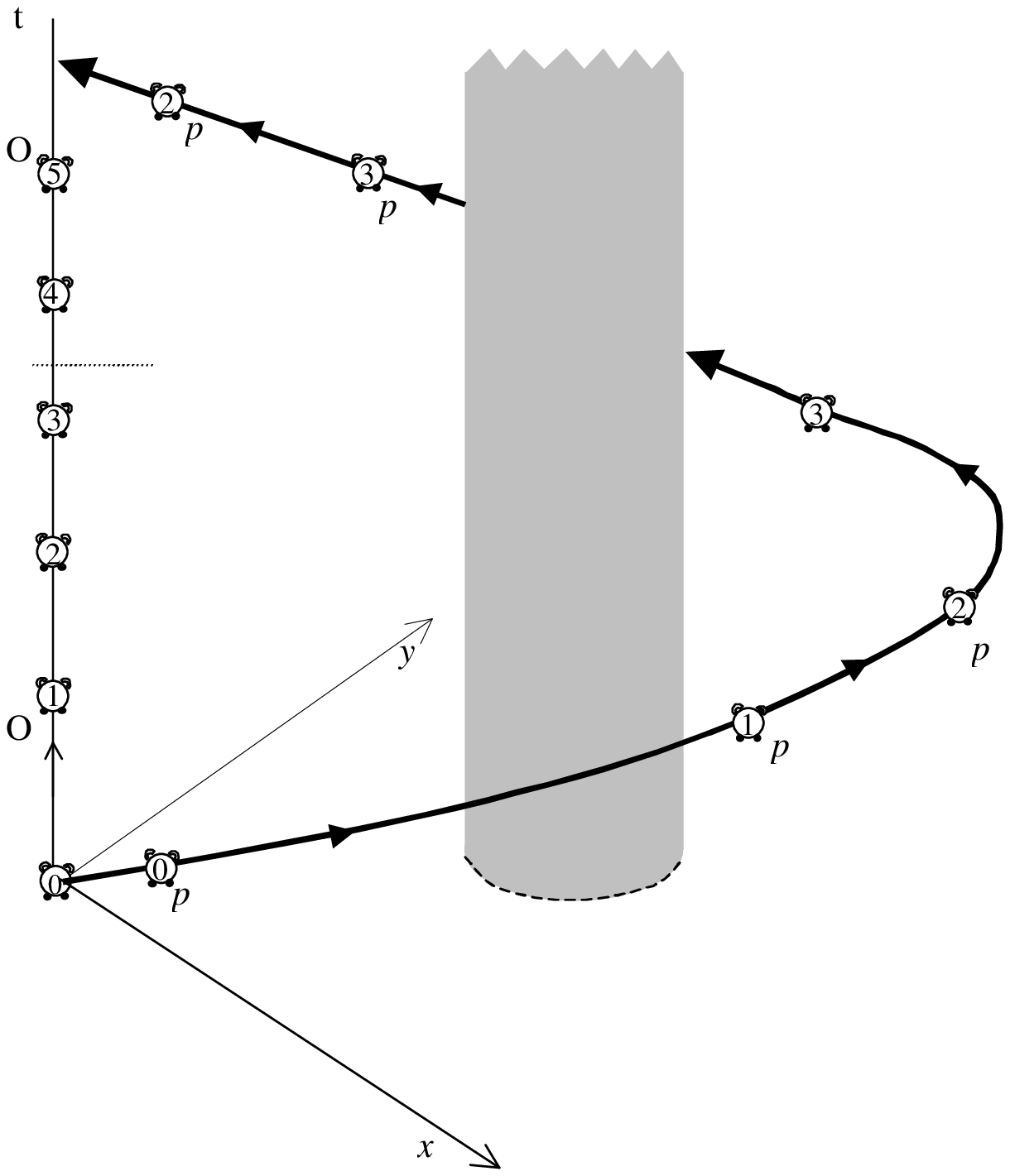}}    }
\put(40,102){\makebox{$\tau$} }
\put(98,125){\makebox{\R} }
\end{picture}
\caption{A flawed attempt to  test time-orientability: An Observer, O,
sends a clock,  P, through \R, it exits \R\ and  reappears counting
backwards.  This  is not  successful because the  clock has  failed to
define a time direction throughout the experiment.}
\label{fig:time1}
\end{figure}

A similar test of the time orientability of \R\ would need a probe
that defined a time direction continuously along a trajectory through
\R\ and back to the observer.  A clock would be a suitable probe, with
the positive time direction defined by increasing times displayed on
the clock. Consider a clock
that enters \R\ at an Observer time $\tau$ and emerges at a later
observer time but counting backwards (see Fig.~\ref{fig:time1}).
it might seem like a direct analogy of the space-orientation experiment, but it is not a demonstration of non-time orientability, because in this experiment, the clock increases in value and then decreases. At some point in the path it attains a maximum reading and at that point it does not define a time direction.

\begin{figure}[ht]
\begin{picture}(150,200)
\put(15,-100){\makebox{\epsfbox{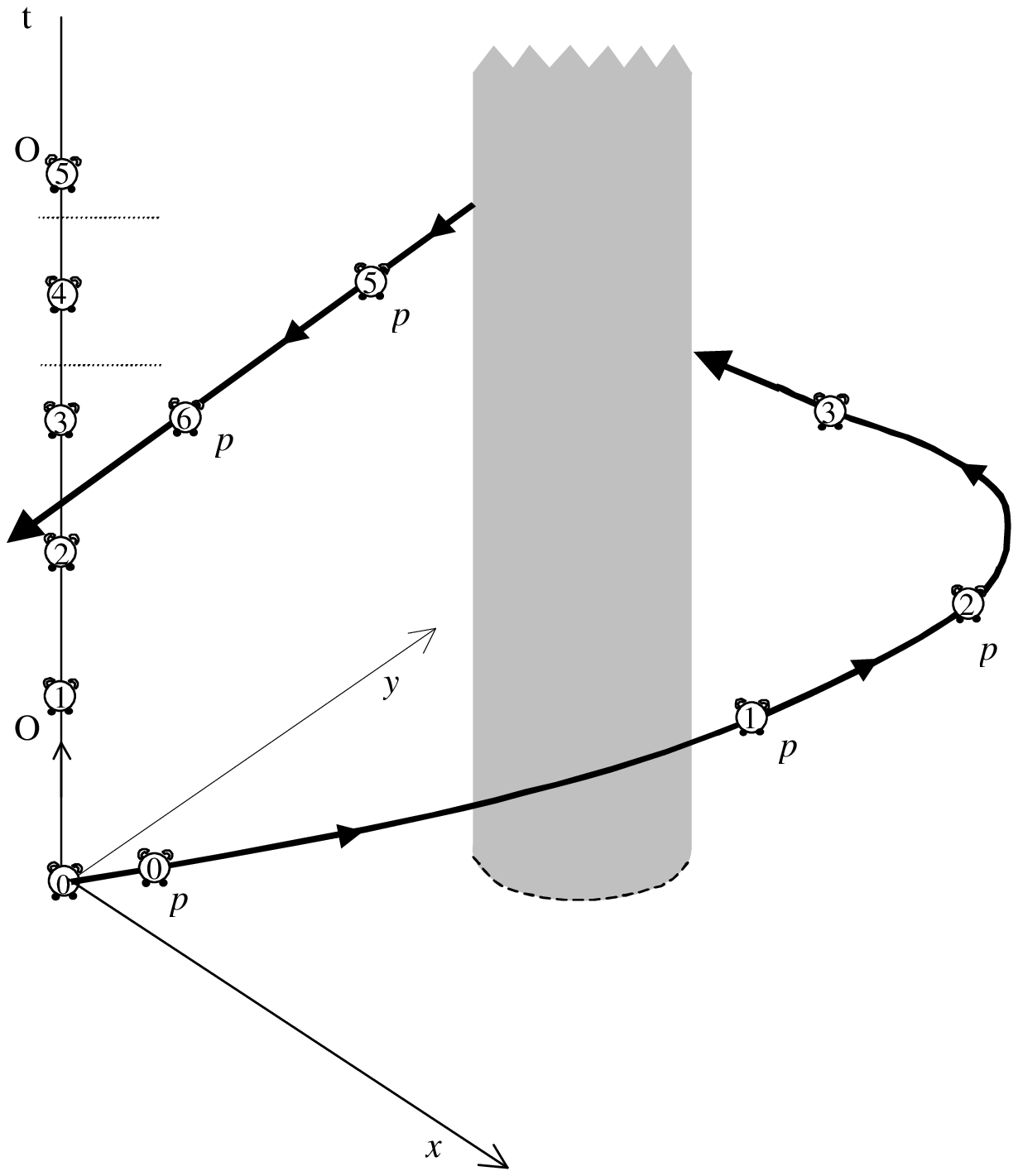}}      }
\put(98,125){\makebox{\R} }
\put(40,121){\makebox{$\rho$} }
\put(40,102){\makebox{$\tau$} }
\end{picture}
\caption{The clock,  P, counts forward continually, it  defines a time
direction  throughout  its path.  The  observer,  O,  sees a  backward
counting clock  and a  forward counting clock  enter \R.  The observer
sees no clocks after observer time $\rho$.}
\label{fig:time2}
\end{figure}

In a true test of time orientability on a region \R\ through which
time cannot be oriented, the clock readings would increase steadily
along the path taken by the clock.  Before entering \R\ the observer
sees the time values increasing on the clock.  When the clock exits at
a time $\rho$ the increasing clock times would be at ever decreasing
values for the observer time. The observer would still see a backwards
counting clock, but only at times before $\rho$ (see
Fig.~\ref{fig:time2}).

This might seem like a clear example of a successful demonstration of
a failure of time orientability.  The author would be sympathetic to
such a view!  However the observer sees the backwards (clock-time
counting backwards) moving clock entering \R\ as well as the forward
counting clock entering \R. At observer times greater than $\rho$ the
observer sees no clock at all - neither inward nor outward moving. So
the observer can interpret the experiment as a clock and an anti-clock
entering \R\ and annihilating each other.

It must be emphasised that this is still a description in terms of
classical physics. The expression {\em anti-clock} is a logical one -
it does not refer to anti-particles in the normal sense - indeed it is
totally independent of the construction of the clock. If you expect
the number of clocks to be preserved, then describing the backward
counting clock as an anti-clock achieves the objective.  Using the
concept of an anti-clock, the experiment has zero clocks most of the
time. Before $\tau$ there is a clock and anti-clock after $\rho$ there
are no clocks.  The alternative is to postulate two rather different
real objects (the clock and the backward counting clock) that both
enter \R\ and vanish. The analogy shares with anti-particles the fact
that both the clock and anti-clock must exist for this to happen.
Unlike the conventional particle-antiparticle annihilation there does
not appear to be any conservation of energy.

To make matters even worse, the experiment is dependent upon the
observer. Since any attempt to stop the backward counting clock from
entering \R\ would be inconsistent with the clock entering and leaving
a time reversing region. So not only is the anti-clock description a
possible alternative description, the time reversing event {\em
requires} the existence of anti-clocks (time reversed clocks) to exist
before the experiment takes place.  Since a clock is normally a
macroscopic object composed of atoms, anti-clocks do not exist in the
sense of clocks composed of antiparticles. Therefore, an apparently
simple test may be physically impossible to carry out.

This highlights a weakness in the way the problem was posed.  Space
orientability is a property of space.  However time orientability can
only be a property of spacetime - a global property. It would appear that the existence or otherwise of a
time reversing region is dependent upon the observer.  A probe and an
anti-probe must both be prepared for a consistent result, in which
case it could be argued that the existence of the time reversing
region depends upon the experiment being performed to test it.

Although  the analysis  is clear,  the interpretation  is by  no means
obvious.  The  following  statements  can  all  be  supported  by  the
arguments above:
\begin{enumerate}
\item  Spacetime   is  not  time   orientable.  Particle  antiparticle
annihilation events are evidence of this.
\item  A  failure  of  time orientability  and  particle  antiparticle
annihilation are indistinguishable.  They are alternative descriptions
of the same phenomena.
\item Time orientability is untestable.
\item  Non  time orientability  cannot  be  an  objective property  of
spacetime  because the  outcome  of  any test  would  depend upon  the
observer.
\end{enumerate}

The conventional  view of the experiment  in Fig.~\ref{fig:time2} is
that  it depicts a  particle antiparticle  annihilation. It is  an
interpretation  that conforms  to  the classical  paradigm -  modelling
Nature in  terms of initial conditions, unique  evolution and observer
independent outcomes.  The classical paradigm  is clearly inconsistent
with a failure of time-orientation.

\section{Orientability and charge}
Manifolds that are not time orientable are enigmatic but possess some fascinating properties. The integral theorems, linking the enclosed charge to the flux on a closed surface have limited applicability. Stokes' theorem requires the enclosed volume to be orientable, while the divergence theorem requires a co-orientation to exist. When the theorems do not apply, it is possible to have the appearance of charge arising from the source free equations because there can be a net surface flux with zero enclosed charge. Sorkin \cite{sorkin} applied the idea using Stokes' theorem to a geon that was not space orientable and found that it could have net magnetic charge. However a 4-geon that is not time orientable does not have a consistently defined normal vector (it is not co-orientable) and the divergence theorem fails, allowing the appearance of net electric charge. A mathematical treatment in arbitrary dimensions is given in \cite{diemer_hadley}.

Working in three dimensions with the vector form of Maxwell's equations, the apparent source of charge is a net flow of flux into or out of a boundary region. A Sphere is the simplest example although all the results apply to an arbitrary compact two dimensional surface that encloses, or appears to enclose a volume of three space. The charge could be either magnetic or electric.
%

Considering first the magnetic charges $Q_m$:
\begin{equation}
Q_m = \oint_{S^2 = \partial V^3} \B.\hat{\textbf{n}}\  \rm{dS}= \int_{V^3} \textbf{Div}. \B \rm{dV} = 0\\
\end{equation}

which neatly relates the flux at a surface to the integrated charge density in the volume enclosed. It shows the apparent charge being due to the sources in the equations. The last step is a simple point by point application of Maxwell's equations. The first step is a global result that requires that the volume, \rm{dV}, is compact and orientable. The integrals also require a metric to be defined. In general the equations do apply and there are no Magnetic charges as a consequence of the vanishing divergence of the magnetic vector field.

A similar treatment for the electric charge gives:
\begin{eqnarray}
Q_e = \oint_{S^2 = \partial V^3} \E.\hat{\textbf{n}}\  \rm{dS} = \int_{V^3} \textbf{Div}. \E \rm{dV} = \int_{V^3} \rho \rm{dV}
\end{eqnarray}

which relates the electric flux at a surface to the integrated charge density in the volume enclosed. As in the case above the volume has to be compact and orientable. But the volume integral also requires a consistent definition of the electric field. However the electric field depends upon the direction of time, and reverses if time is reversed. So it is not necessarily possible to give a global definition of the charge throughout the volume - it will be impossible if the field is non-zero. Consequently, on a non time orientable spacetime, it is possible to have the appearance of electric charges without any sources.

%
The transformation properties of $\E$ are evident from the Lorentz invariant description using the Faraday tensor:%
\begin{equation}\label{eq:F}
    \textbf{F} =
\left(%
\begin{array}{cccc}
     0 & - E_x  & - E_y  & - E_z \\
  E_x & 0    & B_z  & -B_y \\
  E_y & -B_z & 0    & B_x \\
  E_Z & B_y  & -B_x & 0   \\
\end{array}%
\right)
\end{equation}
which shows that the electric field is the space.time component of the tensor and therefore changes its value depending on the direction of the time coordinate. In the absence of a global time direction, the electric field cannot be consistently defined even when F is well defined everywhere. By contrast the magnetic field is the space.space components of the field and does not depend upon the time direction (nor even the space direction). Using the Faraday tensor, Maxwell's equations take the simple form: $d \F =0$ and $d \star \F = \star \J$. The Faraday tensor is a two form and can be integrated using Stokes' theorem, which applies in any k-form, \textbf{$\omega$} and(k+1) dimensional volume, $V$:
\begin{equation}
   \int_{S= \partial V} \textbf{$\omega$} = \int_V d \textbf{$\omega$}
\end{equation}

Stokes' theorem requires $V$ to be a compact oriented manifold. (interestingly it does not require a metric).

Magnetic charge can be defined in terms of the Faraday tensor:

\begin{equation}
Q_m = \int_{S= \partial V} \F  = \int_V d \F = 0
\end{equation}

While a similar application to Stokes' theorem for the dual tensor gives:

\begin{equation}
Q_e = \int_{S= \partial V} \star \F =\int_V d \star \F
\label{eq:qef}
\end{equation}
The RHS evaluates to zero in the absence of any sources ($\J =0$). Although the criteria for Stokes' theorem seem to be met for a manifold that is not time orientable, the star operator is defined in terms of an oriented space-time volume element,$\varepsilon$, by:

\begin{equation}
\F \wedge \star \F = \parallel \F\parallel^2 \varepsilon
\end{equation}
so it is not well-defined if time is not orientable, consequently the volume integral in \ref{eq:qef} is not well defined.

The treatment in terms of $\E$ and $\B$ vector fields is equivalent to that in terms of the Faraday two-form and to other approaches using vector densities. The conclusion is the same, that manifolds that are not time orientable can exhibit net electric charge from the source free equations, while the divergence free magnetic field still implies that there can be no net magnetic charge.

\section{Spin-half transformation properties}

The rotational properties of a non-time orientable manifold are also intriguing. Intrinsic spin is a measure of how an object transforms under a rotation. Trying to apply the concept of a rotation to a manifold is not trivial. A geon or 4-geon is an asymptotically flat manifold with non trivial topology, so we can define a rotation as any transformation with appropriate continuity properties, that matches the normal definition in the asymptotic region of the manifold.

\begin{eqnarray}
 & &\R(\theta )M \to M \\
 & &\R(\theta )\R(\phi )x = \R(\theta  + \phi )x \ \  \forall x \in M \\
 & &\R(0)x = x \ \  \forall x \in M \\
 & &\R(\theta )x \to {\RE } (\theta )x \ \  \text{as}\ |x| \to \infty
\end{eqnarray}

The axis of rotation has been omitted, a thorough treatment requires a definition that is manifestly applicable for the full rotation group, but a restriction to a single axis is adequate for our purposes. The definition above is deliberately as general as possible, allowing a very wide range for mappings to be counted as a rotation.

The definition above may suit mathematicians, but it is not appropriate for a physical rotation, such as when a neutron is rotated in the lab by a magnetic field. The distinguishing feature of a physical rotation is that it is parameterised by time. Time moves from 0 to 1 say and the object is rotated from zero degrees to a specified angle. So we refine our definition to a mapping of spacetime points: $\PR (\theta )(x,0) \to (\R(\theta )x,1)$, plus the conditions above.

%


The mathematical rotation defines a path through space, from the original point $x$ to the rotated point:

\begin{equation}
x \to \R(\theta )x,\ \ \gamma _\theta  (\lambda ) = \{ \R(\lambda \theta )x:\ \ x \in M,\lambda  \in [0,1]\}
\end{equation}

%
Similarly, the physical rotation defines a world line starting at each point in spacetime:
\begin{equation}
(x,0) \to \PR(\theta)(x,0) = (\R(\theta )x,1), \ \ \chi _\theta  (t) = \{ (\R (t \theta) ,t)x:x \in M,t \in [0,1]\}
\end{equation}


But this construction defines a time direction throughout the manifold. It cannot apply to a spacetime that is not time orientable. A modification of the definition is required:

\begin{eqnarray}
  & & \PR (\theta )(x,0) \to (\R(\theta )x,\phi (\theta )(x)) \\
  & & \phi (\theta )(x) \to 1\ \ \text{as}\ ¦x¦ \to \infty
\end{eqnarray}

Where $\phi$ is zero a time direction is not defined by the rotation. The function $\phi$ must be zero over a two dimensional subspace if the manifold is not time orientable. The rotations have two categories of points that do not move:
\begin{description}
  \item[Fixed Points] $(x,0) \to (x,t)$ for example the points on the axis of the rotation.
  \item[Exempt Points] $(x,0) \to (x,0)$ in a sense these points do not participate in the rotation.
\end{description}

Consequently, a physical rotation that is 360 degrees in the asymptotic region will leave some exempt points in the manifold, and it will not be possible to find a 360 degree rotation that is the same as the identity, zero degree rotation. The topology of the rotation group is such that it is possible to find a 720 degree rotation that is an identity.
%
%

The idea has been known for centuries and is modelled by Wheeler's cube in a cube \cite{MTW}, Feynman's scissors trick \cite{feynman_weinberg} and Hartung's tethered rocks \cite{hartung}. A computer animation can be seen on the web \cite{spinhalf}.  What all the demonstrations have in common is a set of fixed points that are not rotated, but they all lack any explanation of how parts of an elementary particle could be anchored in free space. This model of particles as 4-geons provides an explanation for the anchoring and naturally describes particles with spin-half.

It is notable that the requirement for non time orientable manifolds came from the desire to model elementary particles and quantum phenomena. The existence of charge and spin half arise as a natural consequence.

%

%
%

\section{Conclusion}

The results described above apply to any geometric theory of space and time that allows non trivial topology. General Relativity is the established theory of spacetime and satisfies the criteria, but most variations of general relativity would give the same result. If meaningful predictions can be assigned to experiments on acausal manifolds, then they will have the same form as those of quantum theory. The results on non-time-orientable manifolds such as apparent electric charge, spin-half and the logic of propositions may be coincidences. But, at least in principle, general relativity with non-trivial causal structure could explain quantum theory and much more besides. General relativity may be the unified theory that Einstein sought for so long.



\bibliographystyle{ws-procs975x65}

\end{document}